\documentstyle[11pt,newpasp,twoside,epsf]{article}
\def\edcomment#1{\iffalse\marginpar{\raggedright\sl#1\/}\else\relax\fi}
\reversemarginpar

\begin{document}
\title{ Radiative cooling in SPH hydrodynamical simulations of  \\
X-ray clusters}                                                           
\author{R. Valdarnini}
\affil{ SISSA Via Beirut 2-4 34014, Trieste, Italy}

\begin{abstract}
The results from hydrodynamical TREESPH simulations of galaxy clusters are used
to investigate the dependence of the final cluster X-ray properties 
upon the numerical resolution and the assumed star formation models for the 
cooled gas. When cold gas particles are allowed to convert into stars the final
 gas profiles show a well defined core radius and the temperature profiles are
 nearly flat. 
Star formation methods based on a local density
threshold, as in Navarro and White (1993), are shown to give stable 
results. Final X-ray luminosities are found to be numerically stable,
with uncertainties of a factor $\sim 2$.
\end{abstract}

\section{Introduction}

Hydrodynamical simulations have been widely used to predict for different
theoretical frameworks the time evolution of the temperature 
distribution of X-ray clusters. 
With increasing availability of computational power numerical hydrodynamical
simulations have been attempted to model the effects of radiative 
cooling on the gas in the formation and evolution of cluster galaxies
(Katz \& White 1993; Anninos \& Norman 1996; 
Yoshikawa, Jing \& Suto 2000; Pearce et al. 2000; Lewis et al. 2000).
In this contribution I will show the preliminary results that have been 
obtained from a series of 
hydrodynamical SPH simulations of galaxy clusters. The simulations  have 
different numerical resolution and
include the effects on the gas component of radiative losses, star formation 
and energy feedback from SN. 
 Final profiles are compared in order to assess the 
effects of
numerical resolution, or different star formation prescriptions, on the cluster
X-ray variables.

\section{Simulations}
In a previous paper (Valdarnini et al. 1999, hereafter VGB) a 
large set of hydrodynamical simulations was 
used to study the global X-ray cluster morphology and its evolution.
The simulations were run using a TREESPH code with no gas cooling or heating.
I refer to VGB for a detailed description of the simulations.
In Valdarnini (2001) I have used the same cluster sample to study the effects of
 including in the simulations additional physics such as gas cooling and star
 formation.
I will report here the results for the cluster with label $\Lambda$CDM$00$ in 
VGB. This is the most massive cluster ($M_v \simeq 1.5 \cdot 10^{15} 
M_{\odot}$) extracted from a cosmological $\Lambda$CDM N-body 
simulation with size $L=200h^{-1}Mpc$, matter density    
 $\Omega_m=0.3$ and Hubble constant $H_0=70  Km~ sec^{-1}~ Mpc^{-1}$.
In order to check the effects of radiative cooling for this
cluster a set of TREESPH simulations was performed,
 with initial  conditions provided by the cosmological simulation.
The effects of radiative cooling are modelled in these simulations by 
adding to the SPH thermal energy equation an energy-sink term
represented by the radiative cooling function.  
The simulation parameters are:
 number of gas particles, $N_g\sim22,500$; 
  gas particle softening parameter $\varepsilon_g\sim 30 Kpc$; 
gas particle mass, $m_g \sim10^{10} M_{\odot}$; number of dark particles,
$N_d\sim45,000$ (including those in an external shell surrounding the cluster,
cf. VGB).

Allowing the gas to cool radiatively will produce dense clumps of gas
at low temperatures ($\simeq 10^4 \kern 2pt\hbox{}^\circ{\kern -2pt K} $). 
 In these regions the gas
will be thermally unstable and will likely meet the physical 
conditions to form stars.
In TREESPH simulations star formation (SF) processes have been implemented 
using different approaches . 
According to Katz, Weinberg \& Hernquist (1996, hereafter KWH) 
 a gas particle is in a star forming region if the flow
is convergent and the local sound crossing time is larger 
than the dynamical time (i.e. the fluid is Jeans unstable). 
  In a simplified version  
Navarro \& White (1993 , hereafter NW) assume as a sufficient condition 
 that a  gas particle must be in a convergent flow and its density 
exceeds a threshold  ($\rho_g >\rho_{g,c}=7 \cdot 10^{-26} gr cm ^{-3}$).
If a gas particle meets these criteria then it is selected as an eligible 
particle to form stars.
The local star formation rate (SFR) obeys the equation 
 \begin{equation} 
 d\rho_{g}/dt=-c_{\star} \rho_g /\tau_g= -d \rho_{\star} /dt~,
\label{eq:rg}
\end{equation}
where $\rho_g$ is the gas density, $\rho_{\star}$ is the  
star density , $c_{\star}$ is a characteristic dimensionless efficiency 
parameter, $\tau_g$ is the local collapse time and it is the maximum 
of the local cooling time $\tau_c$ and the dynamical time $\tau_d$. 
At each step the probability that a gas particle will form stars in a time 
step $\Delta t$ is compared with a uniform random number. 
 If the test is successful  then a mass fraction
$\varepsilon_{\star}$ of the gas mass is converted into a new 
collisionless particle. This star particle has the position, velocity and
gravitational softening of the original gas particle. 
Typical assumed values are  $\varepsilon_{\star}=1/3 $ and $c_{\star}=0.1$ (KWH).
For the local SFR, NW adopt Eq. 1 with $c_{\star}=1$, $\tau_g=
\tau_d$ and $\varepsilon_{\star}=1/2$ when a gas particle can convert part of its mass into 
a star particle. 
The numerical tests have been performed 
following the  NW prescriptions for selecting gas particles 
which can form stars.
Once a star particle is created it can release energy into the
 surrounding gas through supernova (SN) explosions. 
 All the stars with mass above $ 8 M_{\odot}$ end as a SN, leaving a 
$1.4 M_{\odot}$ remnant.
The SN energy ($\simeq 10^{51} erg$) is released gradually into the gas 
according to the lifetime of stars of different masses. 

\begin{figure*} 
\centerline{\mbox{\epsfysize=4.in\epsffile{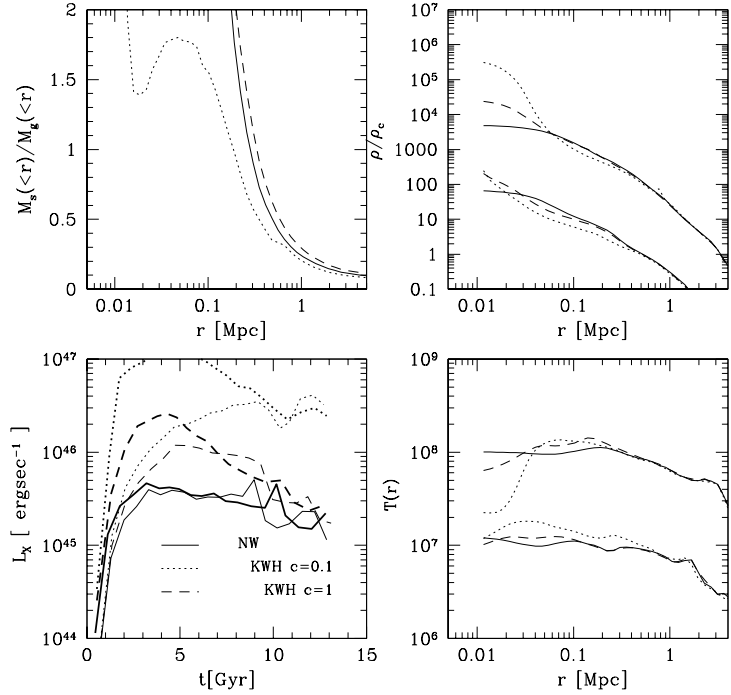}}}
\caption{
Plots showing several cluster properties in simulation runs 
 with different SF prescriptions. 
The continuous line refers to the NW method, the others to KWH with 
different $c_{\star}$ ($c$ in the bottom left panel).
{\it Top left}: ratio of the star mass within the radius $r$ over the 
gas mass within $r$.
 {\it Top right}: final radial density behavior for the gas component.  
The simulation results are compared with the 
corresponding high-resolution runs (see text).  
To facilitate a comparison  
the radial profiles of the high resolution results 
 have been shifted
downward by $10^k$, $k=2$ for densities and $k=1$ for temperatures.
 {\it Bottom right}: Radial temperature profiles.
{\it  Bottom left }: X-ray luminosity versus time, the thick lines 
correspond to the high resolution simulations.}
\end{figure*}

\section{Results}
Three simulations were performed with the same numerical parameters, but using
different SF methods or parameters. 
This in order to test the final dependence of the cluster X-ray variables
on the assumed star formation model for the cooled gas. 
In simulation I cold gas particles have been converted into stars
according to the NW method.
In the other two runs (II and III) the KWH prescription is adopted for
converting gas particles into stars, but with 
 different values of the star formation efficiency parameter
$c_{\star}$:~$0.1$ and $1.$.
As is shown in Fig. 1 the most important differences are found between the KWH  simulations
with different $c_{\star}$. The differences are
dramatic in the final X-ray luminosities, which differ by a factor 
$ \simeq 40$.
The source of this discrepancy relies in the different gas density profiles,
which have substantial differences in the cluster core regions for
$ r {\lower.5ex\hbox{$\; \buildrel < \over \sim \;$}} 100 Kpc$. 
These differences are localized at the cluster center, 
beyond $ r \simeq 100 Kpc$ all the profiles converge, as it is 
shown in the plots of Fig. 1.
The temperature profiles have a peak value of $ \simeq 10^8 
\kern 2pt\hbox{}^\circ{\kern -2pt K}$ at 
$ r \simeq 100Kpc$ and thereafter decline outward by a factor $\sim 2$ 
out to $r_v$. Below $\sim 100Kpc $ the profiles instead show  
large differences. Compared to the NW run the simulation 
with $c_{\star}=0.1$ has gas temperatures which decline inwards by a 
factor $\sim 10$ from $\sim 100Kpc$ down to $r\sim 10Kpc$.
These radial decays follow because of the less efficient conversion
of the cooled gas into stars compared to the NW run.
There is a remarkable agreement for the ratio of cluster mass locked into
 stars to the gas mass, which is $ \simeq 10 \%$  at $r_v$ for all the 
runs considered.
In order to assess the effects of numerical resolution upon 
final results  simulations I, II and III have been run again   
but with a number of particles increased by a factor $ \simeq 3$.
 These simulations will be referred as IH, IIH and IIIH, respectively. 
The simulation results are shown in Fig.~1. For simulations
IH there are not appreciable differences in the radial profiles.
The profiles of 
simulation IIH are instead different from those of run II at 
$r {\lower.5ex\hbox{$\; \buildrel < \over \sim \;$}}
50 Kpc$. The strong drop in $T(r)$ has been removed and 
the gas density profile is much closer to the NW one. 
Simulation IIIH yields final profiles very similar to the ones of 
the parent simulation.
The bottom left panel of Fig.~1 shows that high resolution
runs have final X-ray luminosities which can differ within a factor $\sim 2$
from the parent simulations.

 To summarize, the above results demonstrate that simulations I and III 
 have an adequate numerical resolution to reliably predict 
X-ray cluster properties, such as the X-ray luminosity.
For simulation II ( KWH  with $c_{\star}=0.1$ ) there are large differences 
at the cluster core between the final profiles when the numerical 
resolution is increased.


\begin{references}

\reference{ Anninos, P. \& Norman, M.L. 1996, ApJ, 459, 12 }

\reference{
Katz, N. \& White, S.D.M. 1993, ApJ, 412, 455}

\reference{
Katz, N., Weinberg, D.H. \& Hernquist, L. 1996, ApJS, 105, 19}


\reference{
Lewis, G.F., Babul, A., Katz, N., Quinn, T., Hernquist, L. \& Weinberg, D.H.
2000, ApJ 536, 623}


\reference{
Navarro, J. \& White, S.D.M. 1993, MNRAS, 265, 271 (NW)}
  
\reference{
Navarro, J., Frenk C.S. \& White, S.D.M. 1995, MNRAS, 275, 720}

\reference{
Pearce, F.R., Thomas, P.A., Couchman, H.M.P. \& Edge, A.C. 2000, 
MNRAS 317, 1029}


\reference{
Valdarnini, R., Ghizzardi, S. \& Bonometto, S. 1999, New Astr, 4, 71 (VGB)}

\reference{ Valdarnini, R.  2001, ApJ, in press}

\reference{
Yoshikawa, K., Jing, Y.P. \& Suto, Y. 2000, ApJ, 535, 593 (YJS)}

\end{references}
\end{document}